\documentclass[12pt]{article}
\usepackage{subfigure, graphicx,amsmath,euscript,amssymb,psfrag}
\allowdisplaybreaks

\newcommand{\be}{\begin{equation}}
\newcommand{\ee}{\end{equation}}
\newcommand{\bea}{\begin{eqnarray}}
\newcommand{\eea}{\end{eqnarray}}
\newcommand{\nl}{\nonumber\\}
\newcommand{\order}{{\cal O}}

\newcommand{\sla}[1]{\rlap{\hspace{0.02cm}/}{#1}}

\def\nb{\bar{n}}

\def\nslash{\rlap{\hspace{0.02cm}/}{n}}
\def\nbslash{\rlap{\hspace{0.02cm}/}{\bar n}}

\def\vslash{\rlap{\hspace{0.02cm}/}{v}}

\def\A{{\EuScript A}}
\def\H{{\EuScript H}}

\def\Q{{\EuScript Q}}
\def\X{{\EuScript X}}

\font\title=cmbx10 scaled \magstep3
\font\email=cmtt10

\def\addressphone#1#2{\hbox to \hsize{\@tablebox{#1}\hfil\@tablebox{#2}}}
\def\@tablebox#1{\begin{tabular}[t]{@{}l@{}}#1\end{tabular}}

\begin{document}

\begin{titlepage}

\begin{flushright}
SLAC-PUB-11225\\
{\tt hep-ph/0505129}\\[0.2cm]
May, 2005 \\

\end{flushright}
\vspace{0.7cm}
\begin{center}
\Large\bf 
Heavy-to-light meson form factors at large recoil 
\end{center}

\vspace{0.8cm}
\begin{center}
{\sc Richard~J.~Hill }\\
\vspace{0.7cm}
{\sl Stanford Linear Accelerator Center, Stanford University\\
Stanford, CA 94309, U.S.A. \\
{\email rjh@slac.stanford.edu} } 

\end{center}

\vspace{1.0cm}
\begin{abstract}
\vspace{0.2cm}
\noindent 
Heavy-to-light meson form factors at large
recoil can be described using the same techniques as for hard
exclusive processes involving only light hadrons.  Two competing
mechanisms appear in the large-recoil regime, describing so-called
``soft-overlap'' and ``hard-scattering'' components of the form
factors.  It is shown how existing experimental data from $B$ and $D$
decays constrain the relative size of these components, and how
lattice data can be used to study properties such as the energy
scaling laws obeyed by the individual components.  Symmetry relations
between different form factors ($F_+$, $F_0$ and $F_T$), and between
different heavy initial-state mesons ($B$ and $D$), are derived in the
combined heavy-quark and large-recoil limits, and are shown to
generalize corresponding relations valid at small recoil.  Form factor
parameterizations that are consistent with the large-recoil limit are
discussed.
\end{abstract}
\vspace{1.0cm}

\noindent{\sc PACS: } 12.39.Hg, 12.39.St, 13.20.He, 13.20.Fc 

\noindent{\sc Keywords: } Weak decays, B physics, QCD, form factors, symmetry relations 
\vspace{1.0cm}

\vfill
\end{titlepage}

\section{Introduction \label{sec:introduction} }

Form factors for exclusive heavy-to-light meson transitions at large recoil
energy, such as $B\to\pi l\nu$ with $E_\pi \sim m_B/2$, are an
important ingredient for measurements of the unitarity triangle, and
form the basis for studying more complicated processes such as
radiative $B\to K^*\gamma$ or hadronic $B\to\pi\pi$ decays.  The
description of heavy-meson decays into exclusive final states
containing energetic light hadrons involves multiple energy scales,
and the interplay of perturbative and nonperturbative dynamics.
Simplifications arise upon expanding in powers of the heavy-quark
mass, $m_b$, and the light hadron energy, $E$.  The $1/m_b$ expansion,
when $E$ is small, is implemented by the heavy-quark effective theory
(HQET)~\cite{Neubert:1993mb}, while the $1/E$ expansion, when $m_b$ is
small, is described by well-known methods for hard-exclusive processes
in QCD~\cite{Lepage:1980fj,Efremov:1979qk}.  The simultaneous
expansion for $m_b \sim E \gg \Lambda$, with $\Lambda$ a typical
hadronic scale, requires a merging of these complementary approaches,
and results in the soft-collinear effective theory
(SCET)~\cite{Bauer:2000ew,Bauer:2000yr,Chay:2002vy,Beneke:2002ph,Hill:2002vw}.
This effective field theory description accomplishes the separation of
different energy scales, thus allowing access to the powerful tools of
factorization, to relate different processes to universal hadronic
quantities, and renormalization, to consistently combine perturbative
expansions performed at a high energy scale with the universal
nonperturbative quantities evaluated at a low energy scale.

The focus will be on $B\to P$ transitions, where $P$ is a light
(flavor non-singlet) pseudoscalar meson.  Matrix elements of the
vector and tensor currents are parameterized by the form factors
$F_+$, $F_0$ and $F_T$:
\bea\label{eq:ff_defn}
\langle P(p^\prime)| \bar{q} \gamma^\mu b |\bar{B}(p)\rangle 
&\equiv& F_+(q^2) \left( p^\mu + p^{\prime\mu} - {m_B^2 - m_P^2 \over q^2} q^\mu \right)
+ F_0(q^2) {m_B^2 - m_P^2 \over q^2} q^\mu  \nl
&\equiv& F_+(q^2)  \left( p^\mu + p^{\prime\mu} \right)  + F_-(q^2) q^\mu  \,, \nl
\langle P(p^\prime)| \bar{q} \sigma^{\mu\nu} q_\nu b |\bar{B}(p)\rangle 
&\equiv& {i q^2 F_T(q^2) \over m_B + m_P} \left( 
 p^\mu + p^{\prime\mu}  - {m_B^2 - m_P^2 \over q^2} q^\mu  \right)  \,, 
\eea
with $q \equiv p-p^\prime$.  The vector form factors $F_+$ and $F_0$
are relevant, e.g., in semileptonic $B\to \pi l \nu$, while the tensor
form factor $F_T$ describes, e.g., ``penguin'' amplitudes in $B\to K
l^+ l^-$.  For notational simplicity, results will generally be
written for $B$ decays, with the understanding that similar results
hold also for $D$ decays with $\bar{B}\leftrightarrow D$ ($b
\leftrightarrow c$ at the quark level).

The discussion to follow can be motivated by considering first the
case of small recoil, $E\equiv v \cdot p^\prime = m_P v\cdot v^\prime
\sim \Lambda$, where the velocities are given by $p^\mu \equiv m_B
v^\mu$, $p^{\prime\mu} \equiv m_P v^{\prime\mu}$.  There are two types
of form factor relations that arise in the heavy-quark
limit~\cite{Isgur:1990kf}.  The first type relates different form
factors involving the same initial and final states: at leading order
in $1/m_b$, the form factors appearing in (\ref{eq:ff_defn}) are
related by
\be\label{eq:Fs}
\begin{aligned}
{F_T(q^2) \over m_B + m_P} = {1\over 2m_B}\left[ 
\left(1+ {m_B^2-m_P^2\over q^2}\right)F_+(q^2)  - {m_B^2-m_P^2\over q^2}F_0(q^2) 
\right] \,, 
\end{aligned}
\ee
as follows from (\ref{eq:ff_defn}) upon using $\vslash\, b \approx b$.
The second type relates form factors involving different initial
states, but the same final state (and at the same final-state energy):
\be\label{eq:F_BD}
{F_+^{B\to P}(E) \over F_+^{D\to P}(E)} = \sqrt{m_B\over m_D}\,, \quad
{F_0^{B\to P}(E) \over F_0^{D\to P}(E)} = \sqrt{m_D\over m_B}\,, \quad
{m_D + m_P \over m_B + m_P}{F_T^{B\to P}(E) \over F_T^{D\to P}(E)} = \sqrt{m_D\over m_B}\,, 
\ee
as follows from again using $\vslash\, b \approx b$, and 
the fact that the left-hand sides in (\ref{eq:ff_defn}) scale 
as $\sqrt{m_B}$ ---   
the heavy-quark mass is decoupled from the dynamics via a field redefinition,  
$b(x) = e^{-im_b v\cdot x} h(x) + \dots$, and no other large scales remain.%
\footnote{
The relativistic normalization of states is used throughout. 
}  
HQET formalizes the $1/m_b$ expansion, making explicit the 
mass decoupling at leading power and allowing radiative corrections to be
systematically incorporated.  

Significant modifications should be expected at large recoil for the
symmetry relations (\ref{eq:Fs}) and scaling laws (\ref{eq:F_BD}).
For instance, the dimensionless parameter $v\cdot v^\prime$ can be as
large as $(v\cdot v^\prime)_{\rm max} \approx 6.5$ for semileptonic
$D\to\pi$ transitions, and $(v\cdot v^\prime)_{\rm max} \approx 19$
for $B\to\pi$.  Comparison to a typical dimensionless expansion
parameter of HQET, $m_c/\Lambda \sim 3-4$ or $m_b/\Lambda \sim 10$,
shows that counting $v\cdot v^\prime$ as order unity is likely to lead
to a poor expansion, corresponding to neglect of contributions
involving the new large scale $\Lambda^2 v\cdot v^\prime \sim
E\Lambda$.  The regime of applicability for relations such as
(\ref{eq:Fs}) and (\ref{eq:F_BD}), and the extent to which significant
modifications arise in the kinematic region accessible in $B$ and
$D$ decays, was identified as an important question early in the study
of heavy-quark hadrons~\cite{Isgur:1990kf,Isgur:1990jg}, but has so
far resisted a quantitative understanding.  The present generation of
$B$- and $D$-decay experiments, as well as lattice gauge-theory
simulations, are reaching the level of sensitivity where this question
becomes significant (and answerable).  The appropriate effective field
theory description is now in place to frame this question precisely,
and to interpret the relevant data.

There has been considerable attention in the literature directed at
the description of heavy-to-light form factors at large recoil.
Approaches proceeding in complete analogy with form factors involving
only light mesons suffer from the well-known problems associated with
endpoint singularities~\cite{Szczepaniak:1990dt,Burdman:1992hg}.  In
\cite{Charles:1998dr} it was proposed that the form factors at large
recoil should obey symmetry relations appropriate for the transition
of a static heavy quark into a light energetic quark, thus avoiding
the problem of endpoint singularities by reducing, e.g., all $B\to\pi$
form factors to a single universal (nonperturbative) function.
Symmetry-breaking corrections were investigated in a phenomenological
framework in \cite{Beneke:2000wa}, and the effective-theory description
of these spectator-interaction effects was initiated in \cite{Hill:2002vw}. 
The form factors have since
been studied in more detail using the SCET framework in
\cite{Bauer:2002aj,Beneke:2003pa,Lange:2003pk,Hill:2004if}.  
This paper shows how
these analyses connect to more familiar methods for hard exclusive
processes in QCD, derives new symmetry relations at large recoil that
generalize (\ref{eq:Fs}) and (\ref{eq:F_BD}), and introduces a
convenient parameterization of the form factors for confronting the
experimental and lattice data.

The remainder of the paper is organized as follows.  In
Section~\ref{sec:scet}, the description of form factors in SCET is
outlined, and the close connection between the ideas of SCET and
the description of hard exclusive processes in QCD is demonstrated.
Section~\ref{sec:symmetry} presents the resulting symmetry relations
that are valid at both small and large recoil.  Section~\ref{sec:BK}
introduces a class of parameterizations for the vector $B\to \pi$ form
factors ($F_+$ and $F_0$) which accommodate the new terms appearing at
large recoil.  These new terms require a generalization of
parameterizations often used to present experimental and theoretical
form-factor results.  Section~\ref{sec:expt} considers the
experimental constraints placed on the size of the new terms, and
Section~\ref{sec:theory} compares to recent predictions from lattice QCD
and light-cone QCD sum rules.  Section~\ref{sec:discuss} provides a
concluding discussion.

\section{Form factors in SCET \label{sec:scet} }

To make clear the connection with more familiar ideas from the study
of hard exclusive processes for light hadrons, it is useful to first
consider the description of light meson form factors in the effective
field theory language.  In particular, the matrix elements of the
vector current defining the elastic pion form factor,
\be\label{eq:pipi}
\langle \pi(p^\prime) | \bar{\psi} \gamma^\mu \psi | \pi(p) \rangle 
\equiv \left( p^\mu + p^{\prime\mu} \right) F_\pi(q^2)  \,,
\ee 
and the $\rho-\pi$ transition form factor, 
\be\label{eq:rhopi}
\langle \pi(p^\prime) | \bar{\psi} \gamma^\mu \psi | \rho(p,\eta) \rangle 
\equiv 2i\epsilon^{\mu\nu\rho\sigma}\eta_\nu p_\rho p^\prime_\sigma 
{F_{\rho\pi}(q^2)\over m_\rho + m_\pi } \,, 
\ee
capture all of the essential ingredients required to describe large-recoil 
$B - \pi$ form factors. 

The first task is to decompose $\bar{\psi} \gamma^\mu \psi$ into the
most general effective-theory operator.  Factorization and symmetry
properties for the decay amplitude can then be examined at the
operator level.  In preparation for the discussion of heavy-to-light
form factors, the analysis is done in the rest frame of the
initial-state meson.  To obtain an explicit scale separation, the
momentum modes of the quark and gluon fields $\psi(x)$ and $A_\mu(x)$
are grouped into different momentum regions. A separate
effective-theory field is assigned to each such region, and the
interaction Lagrangian between these effective-theory ``particles'' is
expanded order by order in
$1/E$~\cite{Bauer:2000ew,Bauer:2000yr,Chay:2002vy,Beneke:2002ph,
Hill:2002vw,Becher:2003qh}.
The field decomposition is described in terms of light-cone reference
vectors $n$ and $\nb$, satisfying $n^2=\nb^2=0$ and $n\cdot \nb=2$;
e.g, the default choice is $n^\mu=(1,0,0,1)$ and $\nb^\mu=(1,0,0,-1)$
for an energetic hadron moving in the ${z}$-direction.  A general
momentum can then be expressed as
\be
p^\mu = n\cdot p {\nb^\mu\over 2} + \nb\cdot p {n^\mu\over 2} + p_\perp^\mu \,,
\ee
or more compactly, $p = (n\cdot p\,, \nb\cdot p\,, p_\perp)$.   
The necessary field content of the effective theory 
involves the soft region, with momentum components 
of order $p_s \sim E(\lambda,\lambda,\lambda)$; 
the collinear region, with $p_c \sim E(\lambda^2, 1, \lambda)$; and the 
soft-collinear region, with $p_{sc} \sim E(\lambda^2, \lambda, \lambda^{3/2} )$.%
\footnote{ 
An equivalent ``moving SCET'' description is obtained from
the Lorentz boost $n\cdot p \to \lambda^{-1/2} n\cdot p$, $\nb\cdot p
\to \lambda^{1/2} \nb\cdot p$, under which soft, collinear and
soft-collinear become $\nb$-collinear, $p_{\bar{c}} \sim E^\prime( 1,
\lambda^{\prime 2}, \lambda^\prime)$; $n$-collinear, $p_c \sim
E^\prime( \lambda^{\prime 2}, 1, \lambda^\prime)$; and ultrasoft,
$p_{us} \sim E^\prime( \lambda^{\prime 2}, \lambda^{\prime 2},
\lambda^{\prime 2} )$, respectively~\cite{Becher:2003qh}.  Here
$\lambda^\prime = \lambda^{1/2}$ and $E^\prime \sim \sqrt{E\Lambda}$
are the expansion parameter and energy in the boosted frame.  
}  
Here $\lambda = \Lambda/E \ll 1 $ is a dimensionless expansion parameter.
Fields $\X_c$, $\A_c$ are introduced for collinear particles,
and $\Q_s$, $\A_s$ for soft particles.%
\footnote{
These fields reduce to the ordinary collinear and soft degrees of
freedom $(\xi_c, A_c)$ and $(q_s, A_s)$ in light-cone gauge $\nb\cdot
A_c=0$ and $n\cdot A_s=0$, but in general contain additional gauge
strings to make the operators invariant under soft and collinear gauge
transformations.  
}  
The soft-collinear region, represented by $q_{sc}$ and $A_{sc}$,
describes endpoint configurations of the soft initial-state meson, and
collinear final-state meson, where the $n\cdot p$ and $\nb\cdot p$
components of momentum become atypically small~\cite{Becher:2003qh}.
Sensitivity to this region signals a breakdown of factorization, since
soft-collinear ``messenger'' particles may be exchanged between the
soft and collinear sectors.  Soft-collinear contributions are not
perturbatively calculable, so that demonstrating their absence or
cancellation to all orders in perturbation theory is a necessary
ingredient in establishing factorization for a particular
process~\cite{Becher:2003kh,Lange:2003pk,Becher:2005fg}.  This
cancellation occurs for $F_\pi$ in (\ref{eq:pipi}), but not for
$F_{\rho\pi}$ in (\ref{eq:rhopi}).  Similarly, the $B\to\pi$ form
factors at large recoil contain both a factorizable and a
nonfactorizable piece.

\begin{table}
\begin{center}
\begin{tabular}{|c|c|c|}\hline
 & $d$ & $[\lambda]$ \\[3pt] \hline
$\frac{1}{\nb\cdot \partial_c}\, \bar \X_c \frac{\nb\!\!\!/}{2}
 \Gamma^{\prime} \X_c$ & 2 & 2 \\[3pt]
$\frac{1}{n\cdot \partial_s}\, \bar \Q_s \frac{n\!\!\!/}{2} \Gamma^{\prime}
 \Q_s$ &
2 & 2\\[3pt]
$ \bar \Q_s  \Gamma^{\prime\prime} \Q_s$ & 3 & 3 \\ 
$\!\!{n\!\cdot\! \partial_s}\, \bar \Q_s \frac{\nb\!\!\!/}{2} \Gamma^{\prime}
\Q_s\!\!$ & 4  & 4\\[3pt] 
& $\phantom{\frac{1}{\nb\cdot \partial_c\,n\cdot \partial_s}}$ & \\[5pt]
\hline
\end{tabular} 
\begin{tabular}{|c|c|c|}\hline
 & $d$ & $[\lambda]$ \\[3pt] \hline
$g_\perp^{\mu\nu}$, $\epsilon_\perp^{\mu\nu}$ & 0 & 0 \\[3pt]
$\partial_{c\perp}^\mu$, $\A_{c\perp}^\mu$, $\partial_{s\perp}^\mu$,  
$\A_{s\perp}^\mu$ & 1 & 1 \\[3pt]
$\!\!{n\!\cdot\! \partial_s}\nb\!\cdot\! \partial_s$, 
${n\!\cdot\! \partial_s}\nb\!\cdot\! \A_s\!\!$ & 2 & 2\\[3pt] 
$\!\!{\nb\!\cdot\! \partial_c}n\!\cdot\! \partial_c$, 
${\nb\!\cdot\! \partial_c}\,n\!\cdot\! \A_c\!\!$ & 2 & 2\\[3pt] 
$\frac{1}{\nb\cdot \partial_c\,n\cdot \partial_s}$ & $-2$ & $-1$ \\[3pt]
\hline
\end{tabular}
\end{center}\vspace*{-0.3cm}
\caption{Boost-invariant building blocks for SCET$_{\rm II}$ operators, with
  their dimension $d$ and order $[\lambda]$ in the power expansion. Soft
  derivatives $\partial_s$ can act on any soft field in the operator,
  collinear derivatives $\partial_c$ on any collinear field. 
Here $\Gamma^\prime \in \{1,\gamma_5,\gamma_\perp^\mu\}$, 
$\Gamma^{\prime\prime} \in \Gamma^\prime \cup\{\nb\!\!\!/\,n\!\!\!/,
\gamma_\perp^\mu\gamma_5,\gamma_\perp^{\mu}\gamma_\perp^{\nu}
-\gamma_\perp^{\nu}\gamma_\perp^{\mu}\}$, 
$g_\perp^{\mu\nu} = g^{\mu\nu}-\frac{1}{2}(\nb^\mu n^\nu+n^\mu\nb^\nu)$ 
and $\epsilon_\perp^{\mu\nu}=\frac{1}{2}\epsilon^{\mu\nu\alpha\beta}\,
 {\bar n}_\alpha n_\beta$.  
\label{tab:build}}
\end{table}

At leading order in $1/Q^2$, where $Q^2=-q^2\approx n\cdot p\,
\nb\cdot p^\prime \sim \Lambda E$, the pion form factor
(\ref{eq:pipi}) is given by 
\be\label{eq:Fpi_eff} 
F_\pi = {1\over Q^2}\langle \pi(p^\prime)|[- i\nb\cdot\partial \bar{\psi}\nslash \psi]
| \pi(p) \rangle 
= {1\over Q^2}\langle \pi(p^\prime)|[ in\cdot\partial \bar{\psi}\nbslash \psi] | \pi(p) \rangle \,. 
\ee
In either case, the operator to be represented, $[- i\nb\cdot\partial
\bar{\psi}\nslash \psi]$ or $[ i n\cdot\partial \bar{\psi}\nbslash
\psi]$,
 has dimension four.
The effective-theory representation can be obtained using Table~1 of
\cite{Becher:2005fg}, reproduced here as Table~\ref{tab:build}.%
\footnote{
For a related approach, see \cite{Beneke:2003pa}. 
} 
This table summarizes the building blocks
 which comprise the general effective-theory operators, including
 their mass dimension $d$ and power-counting $[\lambda]$.
We first consider contributions to the matrix element in
(\ref{eq:Fpi_eff}) from ``typical'' momentum configurations of the
initial-state soft pion, and final-state collinear pion, i.e.,
configurations in which none of the partons are in atypical endpoint
momentum regions.  The effective-theory operators must then have the
minimal valence field content $\bar{\X}_c(\dots)\X_c
\bar{\Q}_s(\dots)\Q_s$, in order to mediate the transition of the soft
pion into the energetic collinear pion.  Using the table for $d=4$,
the lowest order in power counting at which this field content can be
realized is $[\lambda]=4$.  From (\ref{eq:Fpi_eff}) it follows
immediately that
\be\label{eq:picount}
F_\pi \sim {1\over Q^2} \,.
\ee 
It can be shown using similar power-counting arguments~\cite{Becher:2005fg} 
that the infrared soft-collinear momentum regions are absent at leading power
from the matrix elements (\ref{eq:Fpi_eff}), and that the resulting
expression takes a factorized form, in terms of a convergent
convolution integral over meson light-cone distribution amplitudes
(LCDAs) and a perturbatively calculable hard-scattering kernel.

The $\rho-\pi$ form factor (\ref{eq:rhopi}) is given by  
\be\label{eq:Frho_eff}
{F_{\rho\pi} \over m_\rho + m_\pi} = {1\over Q^2} (-i)\epsilon_{\perp}^{\mu\nu} \eta^*_\mu 
\langle \pi(p^\prime)| [ \bar{\psi}\gamma_{\perp\nu} \psi] |\rho(p,\eta)\rangle \,, 
\ee
with $\epsilon_\perp^{\mu\nu}$ as in Table~\ref{tab:build}.  The
operator $[ \bar{\psi}\gamma_{\perp\nu} \psi]$ has dimension three,
and from Table~\ref{tab:build}, the leading effective-theory operators
with $d=3$ containing the minimal valence field content
$\bar{\X}_c(\dots)\X_c \bar{\Q}_s(\dots)\Q_s$ have $[\lambda]=4$.
From (\ref{eq:Frho_eff}), it follows immediately that
\be\label{eq:rhocount}
{F_{\rho\pi}\over m_\pi + m_\rho} \sim {1\over Q^4} \,.
\ee
The operators in this case involve at least one occurrence of the
inverse derivative $(\nb\cdot\partial_c n\cdot\partial_s)^{-1}$, as
well as transverse derivatives, extra gluon fields, or the occurrence
of operators such as $\bar \Q_s \Gamma^{\prime\prime} \Q_s$
corresponding to subleading-twist wavefunctions of the initial-state
meson.  Using similar power-counting arguments~\cite{Becher:2005fg}, 
infrared momentum
regions can be shown to contribute at leading power, spoiling
factorization between the soft and collinear sectors.

The mode structure and power counting of SCET, as summarized by the
building blocks in Table~\ref{tab:build}, thus naturally reproduce
the dimensional/helicity counting rules for hard exclusive processes
involving light mesons~\cite{Lepage:1980fj}, e.g.  (\ref{eq:picount})
and (\ref{eq:rhocount}).  The effective theory also allows all-orders
statements concerning factorization to be made; in particular, the
well-known factorizable form of $F_\pi$ at leading order in $1/Q^2$
follows from the uniqueness of the operator with $d=[\lambda]=4$,
together with simple field redefinitions that demonstrate the
cancellation of contributions from infrared momentum regions in the
matrix element of this operator.  The same arguments demonstrate that
$F_{\rho\pi}$ {\it is} sensitive to infrared momentum regions at
leading power, giving rise to the well-known endpoint singularities
that appear in this case~\cite{Chernyak:1983ej}.

The extension to large-recoil heavy-to-light form factors involves one
step in addition to the above analysis.  Here the heavy-quark mass
$m_b$ introduces an additional hard scale into the problem, so that
mapping onto the low-energy theory in this case involves first
integrating out ``hard'' scales of order $\mu^2\sim m_b^2$.  The
resulting description is then independent of the heavy-quark mass, and
the analysis proceeds in direct analogy with the above case, involving
only light hadrons, to integrate out the remaining ``hard-collinear''
scale of order $\mu^2\sim E\Lambda \sim m_b\Lambda$.  The only
difference is the novel (HQET) description of the soft heavy quark,
obtained by replacing the soft Lagrangian by the HQET Lagrangian, and
the soft light-quark field $\Q_s$ by the soft heavy-quark field
$\H_s$.
It should be emphasized that this additional step of integrating out 
the heavy quark can be treated perturbatively, and that   
the remaining low-energy theory consists of precisely the same momentum
modes as for light-meson systems.  

The first step of integrating out hard (but not hard-collinear) scales
is accomplished by matching QCD onto an intermediate effective theory,
denoted SCET$_{\rm I}$.  This intermediate theory describes
hard-collinear and soft fields, with momentum $p_{hc} \sim E(\lambda,
1, \lambda^{1/2})$ and $p_s \sim E(\lambda, \lambda, \lambda)$,
respectively.  Each SCET$_{\rm I}$ operator has a well-defined mass
dimension, and the matching onto the final effective theory, denoted
SCET$_{\rm II}$, is accomplished by using Table~\ref{tab:build} and
the same dimensional arguments as above.  In particular, for the
representation of QCD current operators~\cite{Pirjol:2002km,Hill:2004if},
\be
\bar{q} \Gamma b \to  C^A_i(E,m_b) J^A_i + {1\over 2E} \int_0^1\!\! 
du\, C^B_j(E,m_b,u) J^B_j(u)  + \dots \,, 
\ee
where the SCET$_{\rm I}$ operators of dimension three and four have the 
form
\bea\label{eq:JAB}
J^A_i &=& \bar{\X}_{hc}(0) \Gamma^A_i h(0)  \,, \nl
J^B_j(u) &=& \nb\cdot P\! \int\! {ds\over 2\pi}\, e^{-ius \nb\cdot P} 
\bar{\X}_{hc}(s\nb) \A_{hc\perp\mu}(0) \Gamma_j^{B\mu} h(0) \,.
\eea
Here $\X_{hc}$ and $\A_{hc}$ are hard-collinear quark and gluon fields
and $h$ is the heavy-quark field.  The reference vectors $n$ and $\nb$
are chosen such that $v_\perp=0$.  At leading power the energy is
given by $2E=n\cdot v \nb\cdot P$, where $\nb\cdot P$ is the total
large-component of collinear momentum.  The quantities $u$ and $(1-u)$
in (\ref{eq:JAB}) represent the momentum fractions carried by the
quark and gluon fields, respectively.  Using Table~\ref{tab:build} to
match onto operators with minimal field content $\bar{\X}_c(\dots)\X_c
\bar{\Q}_s(\dots)\H_s$ shows that large-recoil heavy-to-light meson
form factors have two components --- one $A$-type, nonfactorizable,
contribution as in the $\rho-\pi$ form factor, and another $B$-type,
factorizable, contribution as in the pion form factor.  The complete
result at leading power in $1/m_b$ is expressed
as~\cite{Beneke:2000wa,Bauer:2002aj,Beneke:2003pa,Lange:2003pk,Hill:2004if}
\begin{multline}\label{eq:ffdecomp}
F_i^{B\to M}(E) = \sqrt{m_B}\bigg[ 
C^A_{F_i}(E,m_b,\mu)\, \hat{\zeta}_M(E,\mu) \\
+ { 1 \over 2E } \int_0^\infty\!{d\omega\over\omega}{F(\mu)\over 4} \phi_+(\omega,\mu) 
\!\!\int_0^1\!\! du\, f_M(\mu) \phi_M(u,\mu) 
\!\!\int_0^1\!\! du^\prime {\cal J}_\Gamma(u,u^\prime,\ln{2E\omega\over \mu^2},\mu)\, 
C^B_{F_i}(E,m_b,u^\prime,\mu) \bigg] \,, 
\end{multline}
where $M$ represents the light pseudoscalar ($P$) or vector ($V$) final state meson.
Since the $A$-type contribution is nonfactorizable, 
the SCET$_{\rm I}$ matrix element is simply defined by%
\footnote{
$\hat{\zeta}_M$ in (\ref{eq:ffdecomp}) and $\zeta_M$ in \cite{Hill:2004if} are 
related by $\zeta_M = \sqrt{m_B} \hat{\zeta}_M$. 
} 
\be\label{eq:zetadef}
\langle M(p) | \bar{\X}_{hc} \Gamma h  | B(v) \rangle 
 = - 2E\sqrt{m_B} \,\hat{\zeta}_M(E,\mu)\,  {\rm tr}\big[  \overline{{\cal M}}_M(n) \Gamma
 {\cal M}(v) \big] \,, 
\ee
where ${\cal M}(v)$ and ${\cal M}_M(n)$ are spinor wave-functions
appropriate to the heavy-quark and large-energy
limits~\cite{Becher:2005fg}.  The light-cone distribution amplitudes
for the heavy and light mesons are defined by
\bea\label{eq:Fdef}
\langle 0| \bar{\Q}_s(t n)\,{\sla{n}\over 2}\Gamma \H_s(0)
 |B(v) \rangle 
&=& \frac{i F(\mu)}{2}\,\sqrt{m_B}\,{\rm tr}\bigg[{\sla{n}\over 2}
 \Gamma {\cal M}(v) \bigg]
\int_0^\infty
d\omega\,e^{-i \omega t n\cdot v}\,\phi_+(\omega,\mu) \,, \nl
{ \langle M(p) | \bar\X_c(s\nb)\,\Gamma\,\frac{\nb\!\!\!/}{2}\X_c(0)
 |0\rangle }
&=& \frac{i f_{M}(\mu) }{4}\nb\cdot p \,
{\rm tr}\bigg[ \overline{\cal M}_M(n) \Gamma  \bigg] 
 \int_0^1\!
du\, e^{i u s \nb\cdot p } \phi_ M(u,\mu) \,,
\eea
with associated decay constants $F(\mu)$ and $f_M(\mu)$.  Functions
$C^A$ and $C^B$ are matching coefficients for the first matching step
(QCD onto SCET$_{\rm I}$).  The ``jet function'' ${\cal J}_\Gamma$ is
the universal matching coefficient for the factorizable $B$-type
contribution in the second matching step (SCET$_{\rm I}$ onto
SCET$_{\rm II}$), with ${\cal J}_\Gamma={\cal J}_\parallel$ for decays
to pseudoscalar or longitudinally-polarized vector mesons, and ${\cal
J}_\Gamma={\cal J}_\perp$ for decays to transversely-polarized vector
mesons~\cite{Hill:2004if}.

The dependence on energy, heavy-quark mass and renormalization scale
has been made explicit for the various quantities in
(\ref{eq:ffdecomp}).  In particular, the heavy-quark mass dependence
enters the large-recoil heavy-to-light form factors only via two
sources: the overall factor $\sqrt{m_B}$, simply a result of the
relativistic normalization convention for the $B$-meson state; and the
{\it perturbatively calculable} coefficients $C^A$ and $C^B$.  The
energy dependence is also perturbatively calculable for the $B$-type
contribution.  Taking $C^B=-1$, and the tree-level jet function,
\be
{\cal J}_{\parallel}(u,u^\prime)_{\rm tree} = {\cal J}_{\perp}(u,u^\prime)_{\rm tree}
 = - {4\pi C_F \alpha_s\over N}{1\over 2E(1-u)} \delta(u-u^\prime) \,, 
\ee
the quantities%
\footnote{
$\hat{H}_M$ in (\ref{eq:Hdef}) and $H_M$ in \cite{Hill:2004if} are 
related by $H_M = \sqrt{m_B}(m_B/2E)^2 \hat{H}_M$. 
}  
\be\label{eq:Hdef}
\hat{H}_M(E,\mu) \equiv
 { -1 \over 2E } \int_0^\infty\!{d\omega\over\omega}{F(\mu)\over 4} \phi_+(\omega,\mu) 
\int_0^1\! du\, f_M(\mu) \phi_M(u,\mu) 
\int_0^1 du^\prime {\cal J}_\Gamma(u,u^\prime,\ln{2E\omega\over\mu^2},\mu) \,,
\ee
to be considered in detail in the following section, are seen to scale
exactly as $1/E^2$.  Radiative corrections to the jet functions ${\cal
J}_\Gamma$, and to the hard matching coefficients $C^B$, lead to
perturbatively calculable violations of this tree-level scaling law.
The $1/E^2$ law also follows from the heavy-quark mass dependence in
(\ref{eq:ffdecomp}), combined with the SCET power counting, $F_i \sim
\lambda^{3/2}$, and applies to both the $A$-type and $B$-type
contributions.  However, since the nonperturbative function
$\hat{\zeta}_M$ depends on energy, scaling violations for the $A$-type
contributions are not perturbatively calculable.

Both the $A$-type and $B$-type components of the form factors in
(\ref{eq:ffdecomp}) appear at leading power in $1/m_b \sim 1/E$.  The
two components do not mix under renormalization~\cite{Hill:2004if},
and it is then a physically meaningful, and phenomenologically
important, question which, if either, component is dominant.

\section{Symmetry relations in the large-recoil limit \label{sec:symmetry} } 

Given the present uncertainty in the hadronic input parameters
appearing in (\ref{eq:ffdecomp}), it is useful to consider
consequences of this description that are independent of these inputs.
In the following discussion, coefficients $C^A_{F_i}$ and $C^B_{F_i}$
in (\ref{eq:ffdecomp}) will be taken at tree-level.  Higher-order
radiative corrections are small, and their effects are discussed in
Section~\ref{sec:discuss}.  Because $C^B$ is independent of momentum
fraction at tree level, this coefficient can be taken outside of the
convolution integral over $u^\prime$ in (\ref{eq:ffdecomp}), and the
$B$-type contribution is then described by the universal function,
$\hat{H}_M$, introduced in (\ref{eq:Hdef}).

The remainder of the paper focuses on the case $M=P$, i.e., decays
into pseudoscalar final states.  Choosing the normalization of the
$B\to P$ form factors as $F_+$, $(m_B/2E)F_0$ and
$[m_B/(m_B+m_P)]F_T$, the $A$-type coefficients are equal to unity at
tree level: $C^{A(\rm tree)}_{F_+} = C^{A(\rm tree)}_{F_0} = C^{A(\rm
tree)}_{F_T} = 1$.  Also with this normalization, the $B$-type
coefficients are $C^{B(\rm tree)}_{F_+} = 1 - {4E/ m_B}$, $C^{B(\rm
tree)}_{F_0} = -1$, and $C^{B(\rm tree)}_{F_T} = 1$.  From
(\ref{eq:ffdecomp}) and (\ref{eq:Hdef}),
 \bea\label{eq:zetaH} F_+(E) &=& \sqrt{m_B}\bigg[ \hat{\zeta}_P(E) +
  \left({4E\over m_B} - 1\right)\hat{H}_P(E) \bigg] \,, \nl {m_B\over
  2E}F_0(E) &=& \sqrt{m_B}\bigg[ \hat{\zeta}_P(E) + \hat{H}_P(E) \bigg]
  \,, \nl {m_B\over m_B + m_P}F_T(E) &=& \sqrt{m_B}\bigg[
  \hat{\zeta}_P(E) - \hat{H}_P(E) \bigg] \,.  
 \eea

A residual scale dependence is present in the quantities
$\hat{\zeta}_P$ and $\hat{H}_P$, being cancelled by radiative
corrections which have been neglected in the hard-scale coefficients
$C^A$ and $C^B$.  For definiteness, $\hat{\zeta}_P(E) \equiv
\hat{\zeta}_P(E,\mu=2E)$ and $\hat{H}_P(E) \equiv \hat{H}_P(E,\mu=2E)$
in (\ref{eq:zetaH}).  Since the three form factors in (\ref{eq:zetaH})
are described by only two functions, there is one
nontrivial relation between them~\cite{Hill:2004if,Hill:2004rx}:%
 \footnote{ In \cite{Beneke:2002ph} it was shown that with tree-level
 matching there are no contributions from $\order(\lambda^{1/2})$
 SCET$_{\rm I}$ operators that violate the relation
 (\ref{eq:secondclassSCET}).  Since dimensional analysis and
 power-counting~\cite{Beneke:2003pa,Becher:2005fg} shows that no
 $\order(\lambda)$ SCET$_{\rm I}$ operators can match onto SCET$_{\rm
 II}$ operators giving leading-order form-factor contributions, the
 result (\ref{eq:secondclassSCET}) then follows.  Note however that
 the converse is not true - although the symmetry relations for vector
 final states, between form factors $V$ and $A_1$, and between $T_1$
 and $T_2$, receive contributions from $\order(\lambda^{1/2})$
 SCET$_{\rm I}$ operators~\cite{Beneke:2002ph}, these corrections are
 of higher order in the final SCET$_{\rm II}$ power counting, leaving
 exact relations at leading power~\cite{Hill:2004if}.  }
 \be\label{eq:secondclassSCET} {m_B\over m_B+m_P} {q^2\over m_B^2}
 F_T(E) = F_+(E) - F_0(E) \,.  \ee
For comparison, we may rewrite (\ref{eq:Fs}), dropping kinematic
factors quadratic in the light-meson mass, as
 \be\label{eq:secondclassHQET} {m_B\over m_B+m_P}{q^2\over
 m_B^2}F_T(E) = F_+(E) - F_0(E) - {1\over 2}\bigg[ {2E\over m_B}F_+(E)
 - F_0(E) \bigg] \,.
 \ee
Since the extra terms in (\ref{eq:secondclassHQET}) involve factors,
$E/m_B$ or $F_0/F_+$, that are suppressed at small recoil, both
(\ref{eq:secondclassSCET}) and (\ref{eq:secondclassHQET}) are valid
relations at leading power in this regime.  However, at large recoil
the extra terms involving $\hat{H}_P$ are not suppressed, and here
(\ref{eq:secondclassSCET}) is the correct relation.  It is interesting
to note that if the terms involving $\hat{H}_P$ are neglected, then
the relation (\ref{eq:Fs}) (or (\ref{eq:secondclassHQET})) derived at
small recoil is seen to hold in the full kinematic range.  Form
factors describing $B$ decays to vector final states exhibit the same
behavior: the relations derived at small-recoil are equivalent to
relations valid at large recoil, plus hard-scattering corrections.  In
the regime $E\sim m_b \gg \Lambda$, both $\hat{\zeta}_P$ and
$\hat{H}_P$ are of the same order in power counting, and only a
numerical, but not parametric, suppression could justify neglecting
one or the other term.  In fact, when $E\gg m_b \gg \Lambda$ (an
energy regime
 beyond that accessible in $B$ decays), the hard-scattering terms
 dominate, as seen from the fact that the $B$-$\pi$ form factor must
 be described at leading order in $1/E$ in the same way as the
 $\pi$-$\pi$ form factor (cf. (\ref{eq:Fpi_eff}) and (\ref{eq:picount}) above), 
but with an asymmetric $B$-meson
 wavefunction replacing the initial-state pion wavefunction.  It is
 precisely the cross-over regime $E\sim m_b$, where both components
 are of the same order, that is most relevant to experimental studies
 in $B$-decays.

As in the case of small recoil, heavy-quark symmetry may be used to
relate form factors at large recoil for different heavy mesons.  From
(\ref{eq:zetaH}),
 \be\label{eq:BDSCET} 
 {F_-^{B\to P}(E) \over F_-^{D\to P}(E)} =
 \sqrt{m_B\over m_D} \,,\quad {F_0^{B\to P}(E) \over F_0^{D\to P}(E)}
 = \sqrt{m_D\over m_B} \,, \quad 
 {m_D+ m_P\over m_B+m_P}{F_T^{B\to P}(E) \over F_T^{D\to P}(E)} = \sqrt{m_D\over m_B} \,, 
 \ee
with $F_-$ defined in (\ref{eq:ff_defn}).  Both (\ref{eq:F_BD}) and
(\ref{eq:BDSCET}) are valid relations at small recoil, where $(F_- +
F_+)/F_+ \sim 1/m_b$.  However, at large recoil the terms involving
$\hat{H}_P$ enter at leading power, and here (\ref{eq:BDSCET}) gives the
correct relations.

\section{Form factor parameterizations and the large-recoil limit \label{sec:BK} }

The form factor of primary phenomenological interest for $B\to \pi$
decays is $F_+$, since (for massless leptons), it is the only form
factor required to extract $|V_{ub}|$ from the experimental $B\to\pi
l\nu$ rate.  Due to the kinematic constraint $F_+ = F_0$ at $q^2=0$,
it is useful to consider also $F_0$, in order to help constrain
extrapolations of form factor determinations at small recoil
 into the large recoil regime.
It will be convenient to introduce the following normalization and
shape parameters describing these two form factors at large recoil:
 \be\label{eq:params}
  f(0) \equiv F_+(0) \,, \qquad \delta \equiv 1 + {F_-(0) \over
  F_+(0)} \,, \qquad {1\over \beta} \equiv {m_B^2-m_\pi^2 \over
  F_+(0)} {d F_0 \over d q^2 }\bigg|_{q^2=0} \,.
 \ee
From the definition of $F_-$, it follows immediately from
(\ref{eq:params}) that
 \be\label{eq:betaconstraint}
  {m_B^2-m_\pi^2 \over F_+(0)}\left( {d F_+ \over d q^2
  }\bigg|_{q^2=0}
  - {d F_0 \over d q^2 }\bigg|_{q^2=0} \right)
  = 1 - \delta \,.
 \ee
Thus the relative {\it normalization} of $F_+$ and $F_0$ is fixed at
maximum recoil by the kinematic constraint
 \be\label{eq:kinematic}
  F_+(q^2=0) = F_0(q^2=0) \,,
 \ee
while the relative {\it slope} of $F_+$ and $F_0$ is determined by the
quantity $\delta$.  In addition to the quantities (\ref{eq:params})
referring to the large-recoil behavior of the form factors, it is
convenient to introduce the following parameters describing the form
factors at small recoil:
 \be\label{eq:gB}
  {1\over 1-\alpha} \equiv {1\over m_{B^*}^2} {\rm
  Res}_{q^2=m_{B^*}^2} {F_+(q^2)\over F_+(0)} \,, \qquad f(m_B^2) \equiv F_0(m_B^2)
  \,.
 \ee
Note that although the notation anticipates the parameterizations to
be discussed below, the quantities $\alpha$, $\beta$, $\delta$, $f(0)$
and $f(m_B^2)$ have been introduced simply as convenient definitions
for the exact physical quantities appearing on the right-hand sides in
(\ref{eq:params}) and (\ref{eq:gB}).

The parameters $(f(0),\alpha,\beta,\delta)$ are sufficient to describe
the present generation of experimental and lattice form factor data.
Additional shape parameters can be introduced to obtain a
systematically improved form factor parameterization.  A
straightforward approach starts from the dispersive representation,
 \bea\label{eq:dispersive}
  F_+(q^2) &=& { F_+(0)/(1-\alpha) \over 1 - q^2/m_{B^*}^2 }
  + {1\over \pi} \int_{(m_B+m_\pi)^2}^{\infty} dt \, {{\rm Im} F_+(t)
    \over t- q^2 } \,, \nl
  F_0(q^2) &=& {1\over \pi} \int_{(m_B+m_\pi)^2}^{\infty} dt \, {{\rm
  Im} F_0(t) \over t- q^2 } \,,
 \eea
where the $B^*$ pole appears in $F_+$ as a distinct contribution below
the $B\pi$ threshold.  A series of increasingly precise approximations
to the form factors in the semileptonic region, $0 < q^2 <
(m_B-m_\pi)^2$, is obtained by breaking up the integrals in
(\ref{eq:dispersive}), and is given for increasing $N$ by
 \bea\label{eq:ffgeneral}
  F_+ &=& {f(0)/(1-\alpha) \over 1- {q^2\over m_{B^*}^2} }
  + {\rho_1 \over 1- {1\over \gamma_1}{q^2\over m_{B^*}^2} } + \dots
  + {\rho_N \over 1- {1\over \gamma_N}{q^2\over m_{B^*}^2} } \,, \nl
  F_0 &=& {\kappa_1 \over 1- {1\over \beta_1} {q^2\over m_{B^*}^2} }
  + \dots
  + {\kappa_N \over
  1- {1\over \beta_N} {q^2\over m_{B^*}^2} } \,,
 \eea
with parameters constrained by (\ref{eq:betaconstraint}) and
(\ref{eq:kinematic}).  The main focus will be on the case $N=1$, which
can be written~\cite{Becirevic:1999kt}
 \be\label{eq:3param}
  F_+(q^2) = { f(0) \left( 1 - {{1\over \gamma} - \alpha\over
  1-\alpha} {q^2\over m_{B^*}^2} \right) \over \left( 1 - {q^2\over
  m_{B^*}^2} \right) \left( 1 - {1\over \gamma} {q^2\over m_{B^*}^2}
  \right)} \,, \qquad F_0(q^2) = { f(0) \over 1 - {1\over \beta}
  {q^2\over m_{B^*}^2} } \,,
 \ee
where the constraints (\ref{eq:betaconstraint}) and
(\ref{eq:kinematic}) have
been used, and where%
 \footnote{
  Numerical factors $m_{B^*}^2/m_B^2-1$ and $m_\pi^2/m_B^2$ are beyond
  the current level of precision and have been neglected.
 } \be\label{eq:gammainv}
  {1\over \gamma} \equiv 1 - {1-\alpha\over \alpha} \left(
  {1\over\beta} - \delta \right) \,.
 \ee
If the experimental or lattice form factor data can be described by
(\ref{eq:3param}), then the fit parameters
$(f(0),\alpha,\beta,\delta)$
 yield a determination of the physical quantities describing the form
 factors --- at large
recoil on the right-hand sides of (\ref{eq:params}), and at small
recoil on the right-hand sides of (\ref{eq:gB}).

The discussion so far has not required, nor utilized, the
large-recoil, heavy-quark expansion of the form factors.  From
(\ref{eq:zetaH}), the following relations hold at leading order in
$1/m_b$ and $\alpha_s(m_b)$:
 \bea\label{eq:scetparam}
  f(0) &=& \sqrt{m_B} ( \hat{\zeta}_\pi + \hat{H}_\pi )
  \bigg|_{E=m_B/2} + \dots \,, \qquad \delta = {2\hat{H}_\pi \over
  \hat{\zeta}_\pi + \hat{H}_\pi } \bigg|_{E=m_B/2} + \dots \,, \nl
  {1\over\beta} &=& - {d\ \ln ( \hat{\zeta}_\pi + \hat{H}_\pi ) \over
  d\ln E}\bigg|_{E=m_B/2} - 1 + \dots \,.
 \eea
The relations (\ref{eq:scetparam}), between the physical form factors
appearing in (\ref{eq:params}) and the SCET functions
$\hat{\zeta}_\pi$ and $\hat{H}_\pi$, are independent of any
parameterization.  The dependence of the parameters on the
 heavy-quark mass is determined by the scaling laws
$\hat{\zeta}_\pi \sim \hat{H}_\pi \sim 1/E^2$.  In particular, $f(0)
\sim m_b^{-3/2}$, so that $F_+^{B\to\pi}(0)/F_+^{D\to\pi}(0) \approx
(m_D/m_B)^{3/2}$.  
The simple power-counting rules of SCET provide a formal demonstration
of this scaling law, which was justified in \cite{Becirevic:1999kt}
using more qualitative arguments based on QCD sum 
rules~\cite{Chernyak:1990ag,Charles:1998dr}. 
Parameter $\delta$ is $\order(1)$ in the power
counting, and is independent of the heavy-quark mass when scaling
violations are neglected.  Finally, $\beta -1 \sim m_b^{-1}$.  There
are also constraints appearing from the small-recoil regime.  Using
soft-pion relations, it follows that~\cite{Wise:1992hn,Yan:1992gz}
 \be\label{eq:HQETparam}
  {f(0)\over 1-\alpha} = {f_{B^*} g_{B^*B\pi} \over 2 m_{B^*}} \,,
  \qquad f(m_B^2) = {f_B\over f_\pi} \,,
 \ee
where $f_\pi$, $f_B$ and $f_{B^*}$ are decay constants, and
$g_{B^*B\pi}$ is the coupling of the $B$ and $B^*$ mesons to the pion.
If this coupling, and/or the decay constants for the $B$ and $B^*$
mesons were determined precisely, they could be used to place further
constraints on the parameters appearing in (\ref{eq:3param}), or the
more general parameterization
 (\ref{eq:ffgeneral}).
Conversely, to the extent that the data is described by
(\ref{eq:3param}), the resulting fit parameters provide a
determination of $f_{B^*}g_{B^*B\pi}$ and $f_{B}$.  The analysis
in Sections~\ref{sec:expt} and \ref{sec:theory} concentrates on the
region $\alpha < 1$, as required for $f(0)>0$, $g_{B^*B\pi}>0$.  Power
counting in (\ref{eq:HQETparam}) at small recoil, together with the
scaling law $f(0)\sim m_b^{-3/2}$, implies that $\alpha
- 1 \sim m_b^{-1}$ and $f(m_B^2) \sim m_b^{-1/2}$.  The quantity
$\gamma$ in (\ref{eq:gammainv}) is then given up to corrections of
order $1/m_b^2$ by
 \be\label{eq:newgammainv}
  {1\over \gamma} = \alpha + \delta(1-\alpha) \,,
 \ee
yielding the parameterization:
 \be\label{eq:newBK}
  F_+(q^2) = { f(0) \left( 1 - \delta {q^2\over m_{B^*}^2} \right)
  \over \left( 1 - {q^2\over m_{B^*}^2} \right) \left( 1 - \big[
  \alpha + \delta(1-\alpha) \big] {q^2\over m_{B^*}^2} \right)} \,,
  \quad F_0(q^2) = { f(0) \over 1 - {1\over \beta} {q^2\over
  m_{B^*}^2} } \,.
 \ee

Due to the different energy dependence of the coefficients multiplying
$\hat{\zeta}_P$ and $\hat{H}_P$ in (\ref{eq:zetaH}), information on
the parameter $\delta$ describing the relative size of $\hat{\zeta}_P$
and $\hat{H}_P$ can be extracted from the single form factor, $F_+$,
that is most readily
accessible experimentally.%
 \footnote{ The same could not be done from $F_0$, since here
  both terms behave as $E^{-1}$ at large energy.  This fact allows
  $F_0$, but not $F_+$, to be modeled by a single pole.
 }
Several special limits of $F_+$ in (\ref{eq:newBK}) may be noted.  Firstly,
points on the line $\delta=1$ or on the line $\alpha=0$ are equivalent
and correspond to the simple pole model, with a single pole at
$q^2=m_{B^*}^2$.  Secondly, the ``point-at-infinity'', given by
$\alpha\to \infty$, $\delta\to 1$ with $\alpha(1-\delta)$ fixed,
corresponds to a single pole model, with pole at $q^2 =
m_{B^*}^2/[1+\alpha(1-\delta)] $.  If the physical values of the shape
parameters were to lie close to one of these special points, then the
parameter choice $(\gamma, \delta)$, with $\gamma$ from
(\ref{eq:newgammainv}),
 may be more suitable than the choice $(\alpha,\delta)$ for performing
fits; assuming $\alpha > 0$ and $\delta <1$, as indicated by the
data, this will not be the case.  Finally, the axis $\delta=0$
corresponds to the three-parameter Becirevic-Kaidalov (BK)
parameterization~\cite{Becirevic:1999kt}.  It may also be noted that
at small $q^2/m_{B^*}^2$ the shape of $F_+$ is similar to that of
(\ref{eq:newBK}) at $\delta=0$, but with an effective $\alpha_{\rm
eff} = \alpha(1-\delta)$.  Data with sensitivity mostly at small $q^2$
is therefore not easily distinguished from the three-parameter BK
form.  In this situation, since $\alpha=1$ in the heavy-quark limit, a
large deviation of $\alpha_{\rm eff}$ from unity could signal a nonzero
value of $\delta$.  However, with precise enough data, the general
form (\ref{eq:newBK}) can be distinguished from the $\delta=0$ case,
and the parameter $\delta$ measured directly.  The analysis in
Sections~\ref{sec:expt} and \ref{sec:theory} concentrates on the
region $\delta>0$, corresponding to a positive inverse moment of the
$B$-meson LCDA in (\ref{eq:Hdef}).

\section{Experimental Constraints \label{sec:expt} }

\begin{figure}[t]
\begin{center}
\psfrag{x}{$\alpha^{B\pi}$}
\psfrag{y}{$\delta^{B\pi}$}
\includegraphics*[width=20pc, height=20pc]{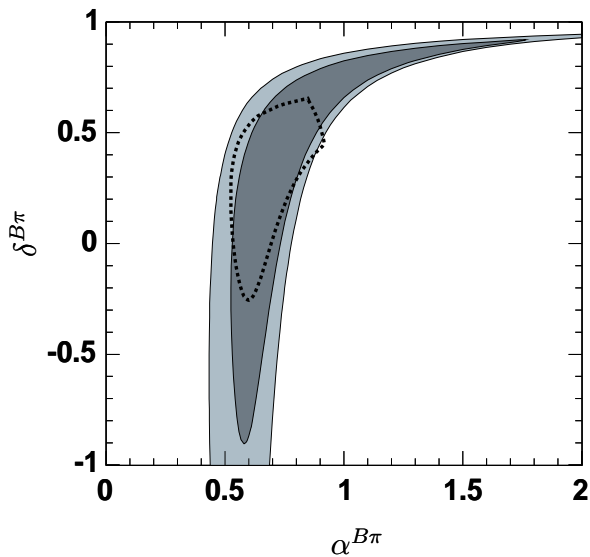}
\caption{$68\%$ (dark) and $90\%$ (light) confidence regions 
for parameters $\alpha$ and $\delta$ as 
determined by fitting $F_+$ in (\ref{eq:newBK}) to 
binned $B\to\pi l \nu$ branching fraction measurements 
in \cite{Athar:2003yg}, \cite{Abe:2004zm} and \cite{babar}. 
Also shown is the boundary of the $68\%$ confidence region (dashed line) 
for the parameterization of $F_+$ in (\ref{eq:3param}), with $\beta^{B\pi}=1.2$.
}
\label{fig:combined}
\end{center}
\end{figure}

Existing experimental data can be used to put significant constraints
on the observables defined in (\ref{eq:params}) and (\ref{eq:gB}).
Figure~\ref{fig:combined} shows the constraints imposed on $\alpha$
and $\delta$ by combined CLEO~\cite{Athar:2003yg} (three $q^2$ bins),
Belle~\cite{Abe:2004zm} (three $q^2$ bins) and BaBar~\cite{babar}
(five $q^2$ bins) $B\to\pi$ branching fraction measurements.  The
contours in Figure~\ref{fig:combined} are obtained from a $\chi^2$ fit
of $F_+$ in (\ref{eq:newBK}) to the data, and correspond to $68\%$
($\Delta \chi^2 = 2.3$) and $90\%$ ($\Delta \chi^2 = 4.6$)
confidence-level regions.  Systematic errors are added to the
statistical errors in quadrature.  
With the exception of \cite{Athar:2003yg}, where
error correlations between different $q^2$ bins are available, 
branching fractions for different bins are assumed uncorrelated, as are
the measurements of different experiments. 
The fit yields
$\alpha^{B\pi}=0.8^{+ 0.5}_{-0.2}$ and
$\delta^{B\pi}=0.6^{+0.3}_{-0.7}$ as the $68\%$ confidence intervals for
the separate parameters.
%~%
% \footnote{
%  The same fit at $\delta=0$ gives $\alpha=0.63^{+0.08}_{-0.10}$,
%  compared to the value $0.60\pm 0.14$ obtained in \cite{babar}.
% }
The simple pole model, corresponding to the boundaries of the plot at
$\alpha=0$ and $\delta=1$, is ruled out decisively by the data
($99.99\%$ level).  However, the single pole model is not ruled out
with high confidence.
The contours in Figure~\ref{fig:combined} thus
extend as fine filaments toward the ``point-at-infinity'' as discussed
after (\ref{eq:newBK}).  If $\delta$ is small, power-suppressed terms
in $\alpha$ and $\beta$ may compete with this parameter in
(\ref{eq:gammainv}).  For comparison, the $68\%$ confidence-level
region obtained from a fit to $F_+$ in the parameterization
(\ref{eq:3param}), before expanding in $\alpha-1$ and $\beta-1$, is
also shown in Figure~\ref{fig:combined}, 
using $\beta^{B\pi}=1.2$~\cite{Shigemitsu:2004ft,Okamoto:2004xg}
(see Section~\ref{sec:theory}), 
and imposing the physical constraint $0 < 1/\gamma < m_{B^*}^2/(m_B+m_\pi)^2$ 
on the position of the effective pole.
With the constraint in place, the unexpanded 
fit yields $1+ 1/\beta^{B\pi} -\delta^{B\pi} = 1.3^{+0.4}_{-0.1}$.%
\footnote{
For a more detailed analysis using 
the general parameterization (\ref{eq:ffgeneral}), see \cite{Becher:2005bg}. 
}  

\begin{figure}[t]
\begin{center}
\psfrag{x}{$\alpha^{D\pi}$}
\psfrag{y}{$\delta^{D\pi}$}
\includegraphics*[width=20pc, height=20pc]{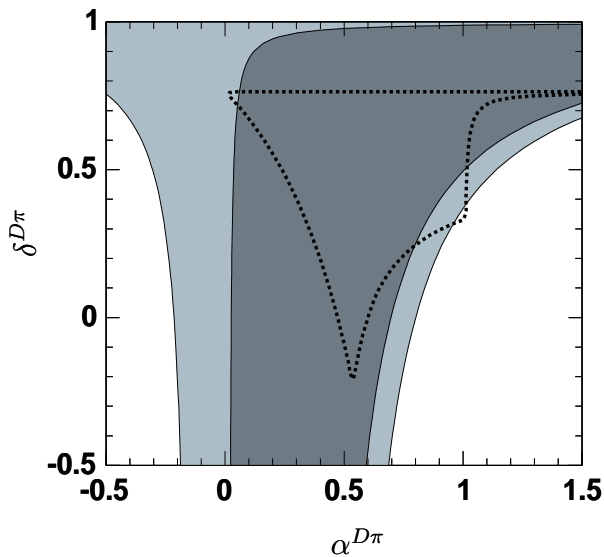}
\caption{
$68\%$ (dark) and $90\%$ (light) confidence regions 
for parameters $\alpha$ and $\delta$ as 
determined by fitting $F_+$ in (\ref{eq:newBK}) (with $m_{B^*}\to m_{D^*}$) 
to $D\to\pi$ data in \cite{Huang:2004fr}.
Also shown is the $68\%$ confidence region (dashed line) using 
the parameterization of $F_+$ in (\ref{eq:3param}) 
before expansion, with $\beta^{D\pi}=1.6$. 
}
\label{fig:cleo_pi}
\end{center}
\end{figure}

Treating the charm mass as sufficiently heavy to perform the
large-recoil/heavy-quark expansion, the same reasoning as led to
(\ref{eq:newBK}) yields a similar parameterization for $D\to\pi$ form
factors, with $m_{D^*}$ replacing $m_{B^*}$.  Under the identification
(\ref{eq:scetparam}), and neglecting scaling violations, 
parameter $\delta$ is independent of the
heavy-quark mass, and is therefore the same for $B$ and $D$ mesons in
the heavy-quark limit.  Figure~\ref{fig:cleo_pi} shows constraints imposed by 
$D\to\pi l\nu$ data from the CLEO collaboration, which measured relative branching
fractions in three $q^2$ bins~\cite{Huang:2004fr}. 
The figure shows $68\%$ and $90\%$ confidence regions 
for $\alpha$ and $\delta$ using the parameterization of $F_+$ in 
(\ref{eq:newBK}), with $m_{B^*} \to m_{D^*}$.
Also shown is the $68\%$ confidence region for the parameterization of $F_+$ in 
(\ref{eq:3param}) before expansion, using 
$\beta^{D\pi}=1.6$~\cite{Aubin:2004ej},%
\footnote{
From the central value for the single-pole fit in 
\cite{Aubin:2004ej}, $\beta^{D\pi}= [m_{D^*}^2/(m_D^2-m_\pi^2)] \times 1.41 \approx 1.6$, 
and $\beta^{D K}= [m_{D_s^*}^2/(m_D^2-m_K^2)] \times 1.31 \approx 1.8$.  
} 
and imposing the constraint $0< 1/\gamma < m_{D^*}^2/(m_D+m_\pi)^2$ on the 
position of the effective pole.
With the constraint in place, the unexpanded fit yields $1+1/\beta^{D\pi}-\delta^{D\pi}=1.1^{+0.6}_{-0.2}$.

\begin{figure}[t]
\begin{center}
\psfrag{x}{$\alpha^{DK}$}
\psfrag{y}{$\delta^{DK}$}
\includegraphics*[width=20pc, height=20pc]{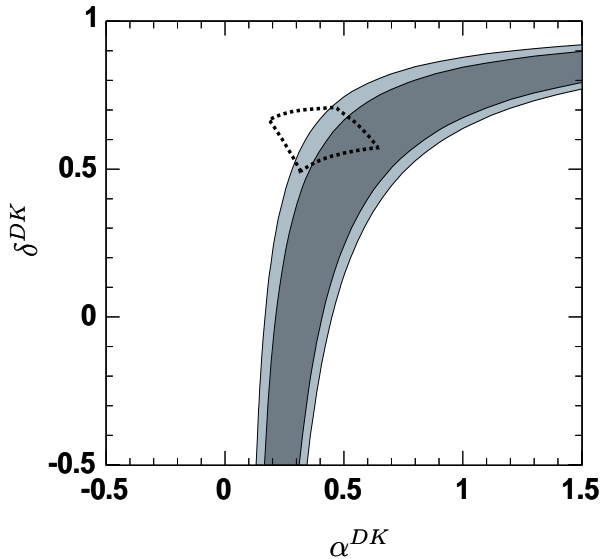}
\caption{
$68\%$ (dark) and $90\%$ (light) confidence regions 
for parameters $\alpha$ and $\delta$ as 
determined by fitting $F_+$ in (\ref{eq:newBK}) (with $m_{B^*}\to m_{D_s^*}$) 
to $D\to K$ data in \cite{Huang:2004fr} and \cite{Link:2004dh}.
Also shown is the $68\%$ confidence region (dashed line) using the parameterization 
of $F_+$ in (\ref{eq:3param}) 
before expansion, with $\beta^{DK}=1.8$. 
}
\label{fig:focus}
\end{center}
\end{figure}

The same analysis can be performed for $D\to K$ form factors, with now
the $D_s^*$ mass being used in (\ref{eq:newBK}).  In the limit of
exact $SU(3)$ flavor symmetry, parameters $\alpha$ and $\delta$ are
the same for this case as for $D\to \pi$ form factors.  The CLEO
collaboration has measured relative branching fractions for $D\to
Kl\nu$ in three $q^2$ bins~\cite{Huang:2004fr}.  The FOCUS
collaboration has extracted the form factor $F_+$ for $D\to K\mu\nu$
decays at nine $q^2$ points~\cite{Link:2004dh}.
Figure~\ref{fig:focus} shows a fit of $F_+$ in (\ref{eq:newBK}), with $m_{B^*}\to m_{D_s^*}$, 
to the combined data, again with $68\%$ and $90\%$ confidence regions.  The
$\chi^2$ fit uses the correlation coefficients from
\cite{Huang:2004fr} and \cite{Link:2004dh} for the respective data,
with the different experiments assumed uncorrelated.  The simple pole
model ($\alpha=0$ or $\delta=1$) is ruled out decisively by the data.
The single pole model ($\alpha\to\infty, \delta\to 1$ with
$\alpha(1-\delta)$ fixed) is not ruled out with high confidence.
Also shown is the $68\%$ confidence region for the 
parameterization of $F_+$ in (\ref{eq:3param})
before expansion, using $\beta^{DK}=1.8$~\cite{Aubin:2004ej},
 and
 imposing the physical constraint $0< 1/\gamma < m_{D_s^*}^2/(m_D+m_K)^2$ on the
position of the effective pole.
With the constraint in place, the unexpanded 
fit yields $1+1/\beta^{D K} -\delta^{D K} = 0.91^{+0.12}_{-0.05}$.
A direct measurement of the quantity $\delta$ (as defined in (\ref{eq:params}))
in \cite{Link:2004dh} yields  $\delta^{DK} = -0.7 \pm 1.5 \pm 0.3$.

\section{Comparison to Theoretical Predictions \label{sec:theory}} 

\begin{figure}[t]
\begin{center}
\psfrag{x}{$\alpha^{B\pi}$}
\psfrag{y}{$\delta^{B\pi}$}
\includegraphics*[width=20pc, height=20pc]{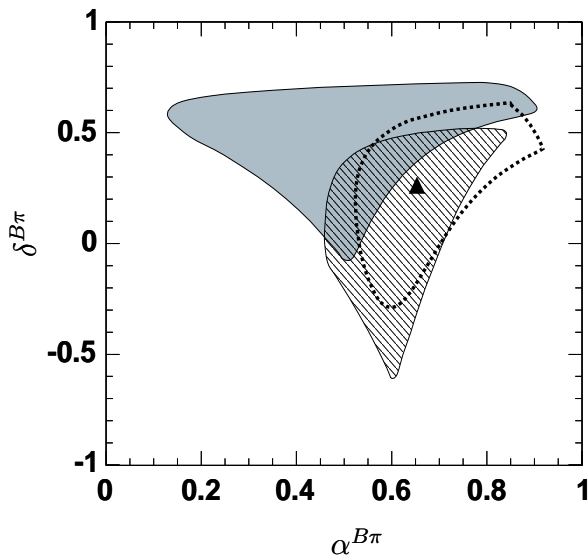}
\caption{
Theoretical constraints on form-factor shape parameters. 
Shown are $68\%$ confidence regions for parameters $\alpha$ and $\delta$ for $B\to\pi$ 
as determined by fitting (\ref{eq:3param}) to 
unquenched lattice QCD  in \cite{Shigemitsu:2004ft} (light solid)
and \cite{Okamoto:2004xg} (hatched).
The triangle indicates the central value from light-cone sum rules in \cite{Ball:2004hn}.
Superimposed is the $68\%$ confidence region  (dashed line)
 from experimental data (see Figure~\ref{fig:combined}). 
}
\label{fig:theory}
\end{center}
\end{figure}

Explicit theoretical form-factor calculations are important, most
notably for supplying the overall normalization necessary to extract
weak-interaction parameters ($|V_{ub}|$) from experimental data
($B\to\pi l\nu$).  The form factor shape can be tested independently
of this normalization, and also contains important information
relating to other processes.  Figure~\ref{fig:theory} shows allowed
parameter regions obtained by fitting recent unquenched $B\to\pi$
lattice data for $F_+$ and $F_0$, 
from \cite{Shigemitsu:2004ft} and \cite{Okamoto:2004xg},
to (\ref{eq:3param}), imposing the physical constraint 
$0 < 1/\gamma <m_{B^*}^2/(m_B+m_\pi)^2$ and 
$0 < 1/\beta < m_{B^*}^2/(m_B+m_\pi)^2$
on the positions of the effective poles. 
Also shown is the central value for the
light-cone sum rule determination from
\cite{Ball:2004hn}.  Superimposed is the region preferred by the
experimental data, as in Figure~\ref{fig:combined}.  It may be noted
that the lattice determination of $F_+(q^2)$ in \cite{Okamoto:2004xg}
employs the BK parameterization, (\ref{eq:newBK}) with $\delta=0$, to
interpolate and extrapolate the data points at varying light-quark
mass to fixed energy prior to performing the chiral extrapolation to
the physical light-quark mass.  Similarly, a single pole model is used
in \cite{Shigemitsu:2004ft} to interpolate to fixed energy before
chiral extrapolation.  Achieving greater precision warrants further
investigation into whether the form assumed for this extrapolation
biases the resulting chirally-extrapolated form factors.  Also, as a
result of fitting to a smooth curve prior to chiral extrapolation, the
data points for $F_{+,0}(q^2)$ from \cite{Shigemitsu:2004ft} and
\cite{Okamoto:2004xg} lie on a smooth curve, introducing significant
correlations between different $q^2$ values.  The $\chi^2$ fit
employed in Figure~\ref{fig:theory} assumes uncorrelated errors, with
statistical and systematic errors added in quadrature.

\begin{figure}[t]
\begin{center}
\subfigure[]{\label{fig:slopez}
\psfrag{x}{$\alpha^{B\pi}$}
\psfrag{y}{\hspace{-0.5cm}$- {d \ln\hat{\zeta}_\pi \over d \ln E}|_{E=m_B/2}$}
\includegraphics*[width=19pc, height=15pc]{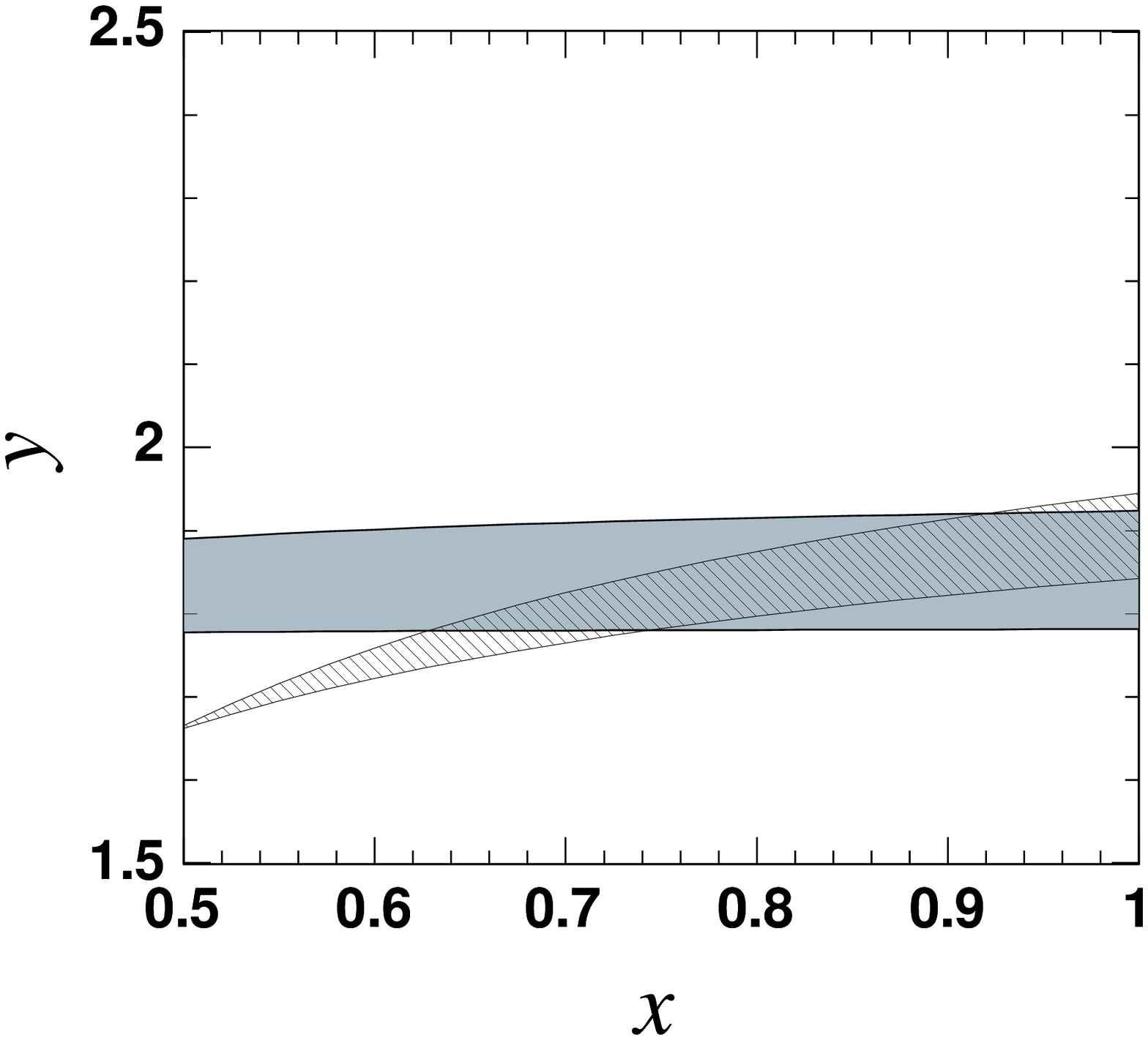}
}
\subfigure[]{\label{fig:slopeh}
\psfrag{x}{$\alpha^{B\pi}$}
\psfrag{y}{\hspace{-0.5cm}$- {d \ln\hat{H}_\pi \over d \ln E}\big|_{E=m_B/2}$}
\includegraphics*[width=19pc, height=15pc]{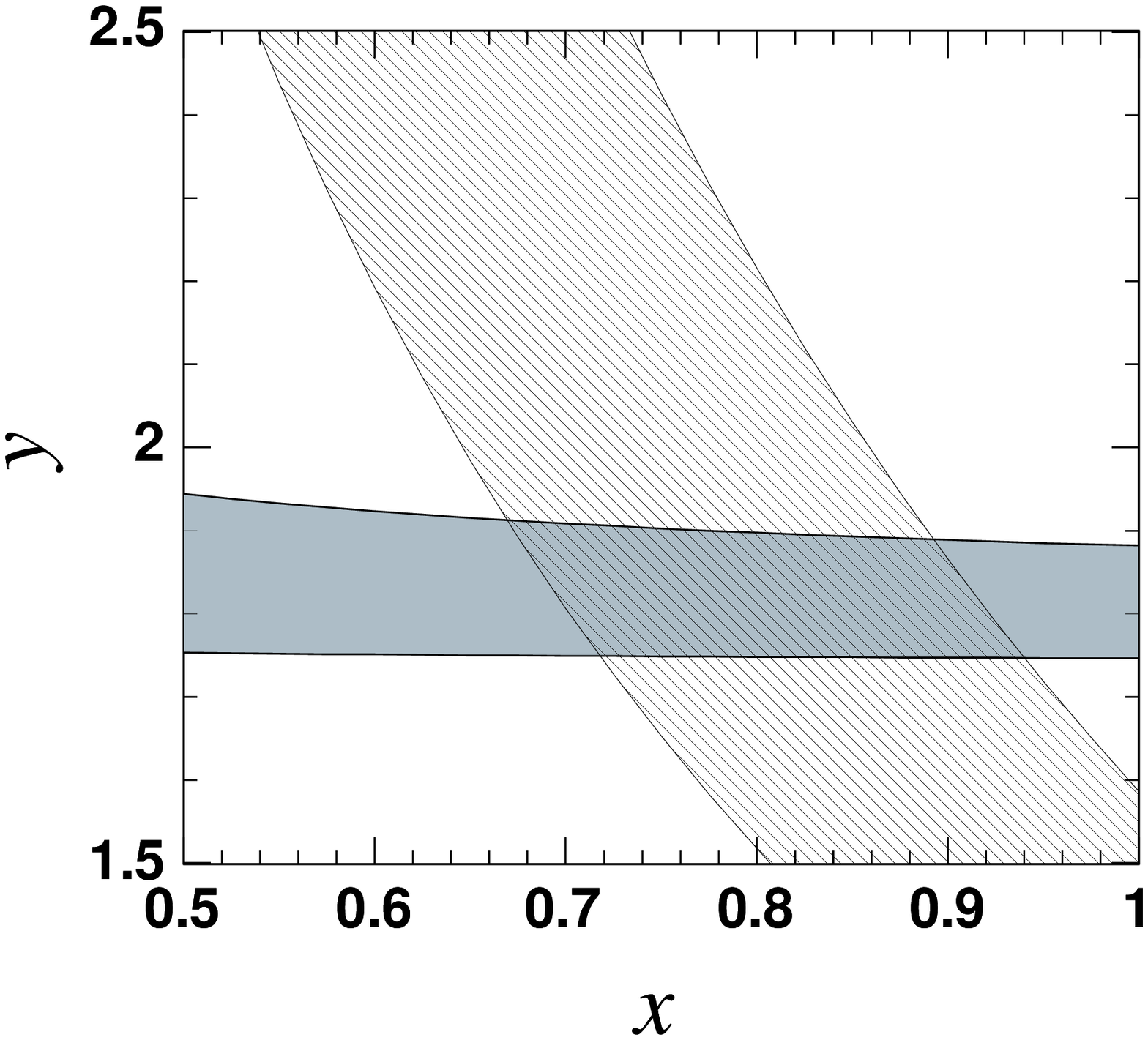}
}
\caption{
Slope of $\hat{\zeta}_\pi$ (\ref{fig:slopez}) 
and $\hat{H}_\pi$ (\ref{fig:slopeh}) at maximum recoil, 
with $F_+$ and $F_0$ given by (\ref{eq:3param}), 
as a function of $\alpha$.   The hatched region is for $\delta=0.2$, 
and the solid region for $\delta=0.7$.  The vertical range within each region
corresponds to varying $\beta$ from  $\beta=1.1$ to $\beta=1.3$.  
}
\label{fig:slope}
\end{center}
\end{figure}

Under the assumption of the dimensional scaling laws
$\hat{\zeta}_\pi\sim 1/E^2$ and $\hat{H}_\pi \sim 1/E^2$, the
identification (\ref{eq:scetparam}) allows the relative size of
$\hat{\zeta}_\pi$ and $\hat{H}_\pi$ to be probed by measuring the
single form factor $F_+$, using (\ref{eq:zetaH}).  By measuring both
$F_+$ and $F_0$, it is possible to isolate the $\hat{\zeta}_\pi$ and
$\hat{H}_\pi$ components directly, and to test this scaling law.
 With the form factors
parameterized according to (\ref{eq:3param}),
small deviations of $\alpha$ and $\beta$ from unity allow for
violations of the exact $1/E^2$ scaling that is recovered when
$\alpha=\beta=1$ (for any value of $\delta$).  Figure~\ref{fig:slope}
shows the range of slopes for typical parameter values.  Parameter
$\beta$ is varied between $1.1$ and $1.3$; this is consistent with the
lattice values%
 \footnote{
  The fits in \cite{Shigemitsu:2004ft} and \cite{Okamoto:2004xg}
  assumed $\delta=0$, but the value of $\beta$ is not significantly
  changed by allowing $\delta\ne 0$.
 } $1.18(5)$ from \cite{Shigemitsu:2004ft} and $1.18(5)$
from \cite{Okamoto:2004xg}, and also with the light-cone sum rule
result $1.20$ from \cite{Ball:2004hn}.   The magnitude of the
slope for $\hat{\zeta}_\pi$ in Figure~\ref{fig:slopez} is slightly
below $2$ for all parameter values, whereas for $\hat{H}_\pi$ the
result shown in Figure~\ref{fig:slopeh} depends more sensitively on
the value of $\delta$.  Small deviations from the $1/E^2$
dimensional scaling law lend confidence to the large-recoil
expansion.

\section{Discussion \label{sec:discuss} }

Form factors at large recoil energy are essential to the study of
heavy meson decays to exclusive final states.  The additional energy
scale provided by the heavy quark complicates the description of these
processes relative to the case of exclusive processes involving only
light hadrons.  However, in the minimal effective theory (SCET$_{\rm
I}$) obtained after integrating out the heavy-quark mass, the
description follows that of the light-hadron case, with a novel HQET
field replacing one of the light-quark fields.  The problem can be
made to look more symmetric by boosting to the Breit frame for the
light degrees of freedom.  In continuum field theory the SCET and
``moving SCET'' descriptions are of course equivalent; however, in
lattice simulations, the maximum light-meson energy in the Breit frame, $E^\prime
\sim \sqrt{E\Lambda}$, is much smaller than that in the rest frame of
the initial-state heavy meson.  The resulting discretization requires
far fewer lattice sites to obtain a given accuracy, and can lead to
much more efficient simulations.  This philosophy lies behind the idea
of ``moving nonrelatvistic QCD'' and related approaches~\cite{Foley:2002qv}, which
may allow direct simulations of form factors over most or all of the
full kinematic range in $B\to\pi l\nu$.

Knowledge of the heavy-quark mass dependence of the form factors, and
of the relations between different form factors, provides a valuable
handle that can be used to test lattice or other theoretical
calculations.  For example, if $F_T$ were calculated on the lattice,
(\ref{eq:secondclassSCET}) is a model-independent relation that must
be satisfied throughout the entire kinematic range.  If the
hard-scattering contributions are significant, then there is a
nontrivial modification at large recoil compared to the corresponding
HQET relation (\ref{eq:Fs}). 

Symmetry relations between form factors for $B$ and $D$
mesons near the kinematic endpoint for $D$ decay can provide a
normalization for the $B$-decay form factors.  Relations
(\ref{eq:BDSCET}) are valid throughout the entire kinematic range;
again, if the hard-scattering contributions are significant, there is
a nontrivial modification from the corresponding HQET relations
(\ref{eq:F_BD}).  
 The analysis of $B$ decays to
vector final states reveals similar modifications at large recoil to
the HQET scaling laws and symmetry relations; a preliminary discussion
was given in \cite{Hill:2004rx}, and a more in-depth analysis is left
for future work.
Precision measurements would require a detailed
study of power corrections, in particular an analysis of the manner in
which HQET power corrections~\cite{Burdman:1993es} merge with the SCET
description; this topic is beyond the scope of the present work.
However, it should be emphasized that, even at leading order in the heavy-quark expansion, 
no formal relation exists
between form factors $F_+^{B\to\pi}$ and $F_+^{D\to\pi}$ near maximum 
recoil in the $D\to\pi$ system.

Neglecting scaling
violations (and power corrections), the large-recoil scaling laws imply
$F_+^{B\to\pi}(q^2=0) = (m_D/m_B)^{3/2} F_+^{D\to\pi}(q^2=0)$.  As
emphasized in Section~\ref{sec:scet}, the energy dependence of the
hard-scattering part of the form factors is calculable; for this part,
a full renormalization-group analysis including first-order radiative
corrections was performed in \cite{Hill:2004if}, using a model
$B$-meson wavefunction.  Scaling violations were found to give a small
additional suppression for increasing energy relative to the $1/E^2$
tree-level scaling.  For the soft-overlap part of the form factors,
however, the energy dependence is not perturbatively calculable.
Scaling violations for such nonfactorizable quantities provide an
interesting window on nonperturbative QCD dynamics.  Although for
practical purposes, this interesting dynamics can fortunately be
avoided through the use of symmetry relations, further direct analysis
of $\hat{\zeta}_\pi$ is warranted.

The description in Section~\ref{sec:symmetry} used the tree-level
approximation for the hard-scale matching coefficients in the
effective theory.  Like the HQET relations (\ref{eq:Fs}) and
(\ref{eq:F_BD}), the SCET form factor relations
(\ref{eq:secondclassSCET}) and (\ref{eq:BDSCET}) receive corrections
at $\order(\alpha_s)$, arising as nontrivial matching coefficients of
QCD onto the effective theory at the hard matching scale.  These
radiative corrections have been calculated to first order in
$\alpha_s$, for $C^A$ in \cite{Bauer:2000yr,Beneke:2004rc}, and for
$C^B$ in \cite{Beneke:2004rc,Becher:2004kk}.  The corrections could be
taken into account trivially for the $A$-type terms; however, since
the modifications are $\lesssim 5\%$ for all $B\to\pi$ form 
factors~\cite{Becher:2004kk}, they can be safely neglected at the
current level of precision.  Corrections to $C^B$ are more difficult
to quantify, since beyond tree level these coefficients are
momentum-fraction dependent, and so knowledge of the shape of the
meson wavefunctions becomes necessary.  However, these corrections are
$\lesssim 20\%$ for all $B\to\pi$ form 
factors~\cite{Becher:2004kk}; if, as the data
indicate, the hard-scattering terms ($\hat{H}$) are significantly
smaller than the soft-overlap terms ($\hat{\zeta}$), then the effect
on the overall form factors is much smaller.  Until evidence of the
hard-scattering terms is first seen unambiguously and their properties
can be studied further, this approximation is sufficient.

The parameterization (\ref{eq:newBK}) provides a generalization of
several forms commonly used in studying form factors for
heavy-to-light transitions such as $B\to\pi$.  
It is the most general form of 
$F_+$ with a pole at $q^2=m_{B^*}^2$ and one additional effective pole. 
The simple pole (one pole at $m_{B^*}^2$) and single pole 
(one pole, not necessarily at $m_{B^*}^2$) models are contained as
special cases.  
Given that the lattice and experimental data cannot yet resolve 
more than a single effective pole, 
branching fraction fits and the resulting  $V_{ub}$  determinations
should be very insensitive to the inclusion of even more terms in (\ref{eq:ffgeneral}). 
In particular, with the asymptotic behavior $F_+(t)\sim 1/t$ at large $t$, the 
dispersive integral in (\ref{eq:dispersive}) is absolutely convergent. 
This bounds the magnitude of residues $\rho_i$ in (\ref{eq:ffgeneral}), 
and prevents contributions from arbitrarily large $t$.   
A more formal study of the convergence properties of the sequence of
parameterizations  (\ref{eq:ffgeneral}) will be taken up elsewhere~\cite{Becher:2005bg}. 
Using the identification (\ref{eq:scetparam}), 
it follows that 
specialization to the case $\delta=0$ in (\ref{eq:newBK}), as in \cite{Becirevic:1999kt}, 
corresponds to neglect of the hard-scattering component of the form factor, 
and is justified only to the extent that this component is shown 
to be small.  
Similar
parameterizations can be used in $D\to\pi$, and in $D\to K$ decays,
with the $D^*$, and $D_s^*$ pole replacing the $B^*$ pole.  
%
%It remains
%to be seen whether the hard-scattering terms can be uncovered already
%in the energy range accessible in $D$ decays.  
%
It would be especially
interesting to fix the value of $\delta = 1 + F_-(0)/F_+(0)$ in
$D\to\pi$ decays.  From the identifications (\ref{eq:scetparam}), this
quantity is independent of the heavy-quark mass at leading power, and
neglecting scaling violations.  Since $\delta$ is at most a
slowly-varying function of the heavy-quark mass, determination of its
value for $D\to\pi$ would give an important indication of its size for
$B\to\pi$.

A related decomposition of the dispersive integral (\ref{eq:dispersive}) 
was studied in \cite{Burdman:1996kr}, where a parameterization in terms of
explicit resonance and continuum contributions was put forward. 
The semileptonic data can now be used to test such models --- e.g.,
$B\to \pi$ (Figure~\ref{fig:combined}) and 
$D\to K$ (Figure~\ref{fig:focus})  data definitively resolve contributions
other than the $B^*$ and $D_s^*$ pole terms, respectively, 
and $D\to\pi$ (Figure~\ref{fig:cleo_pi}) data favors contributions in addition
to the $D^*$ pole, as indicated by $\alpha > 0$.   

A detailed analysis of heavy-to-light form factors provides the basis
for more complicated radiative and hadronic $B$ decays.  For instance,
at leading order in $1/m_b$, $B\to \pi\pi$ decays can be related via
factorization theorems to $B\to\pi$ form factors, plus hard-scattering
corrections~\cite{Beneke:1999br,Beneke:2000ry}.  Written in SCET
language, the leading-order description is in terms of the same
functions $\hat{\zeta}_\pi$ and $\hat{H}_\pi$ appearing in the form
factors~\cite{Bauer:2004tj}.  Knowledge of the parameter $\delta$ in
(\ref{eq:zetaH}) should help in understanding these more complicated
processes, where values ranging from $\delta \sim 0.1
-0.5$~\cite{Beneke:2003zv,Beneke:2004bn} to $\delta\sim 1.1 -
1.4$~\cite{Bauer:2004tj,Bauer:2005wb} have been taken as
phenomenological input or suggested from fits to the $B\to\pi\pi$
data.  The size and nature of power corrections also warrants further
investigation~\cite{Feldmann:2004mg}.

More generally, the value of $\delta$ should help in deciding between
different schools of thought that have emerged to describe $B\to\pi$
form factors.  The first of these may be conveniently labelled as the
``soft-overlap dominance'' school, where $\delta$ is small, and the
hard-scattering terms at large recoil give small corrections to the
symmetry relations derived at small recoil; the second,
``hard-scattering dominance'', school of thought, where $\delta$ is
large, assumes that the $B\to\pi$ transition can be treated in much
the same way as for light-meson form factors, where hard-scattering
terms are dominant and endpoint contributions are suppressed.
Light-cone sum rules generally belong to the first school, where the
hard-scattering terms appear as a radiative
correction~\cite{Bagan:1997bp,Ball:1998tj,DeFazio:2005dx}.  Some care is required in
categorizing various approaches in the second school, due to different
terminologies.  In SCET, it is natural to identify the
``soft-overlap'' contribution with the ``nonfactorizable''
$\hat{\zeta}$, arising from the ``$A$-type'' SCET$_{\rm I}$ current,
and satisfying ``spin-symmetric'' relations (\ref{eq:zetadef})
appropriate for this leading-order current operator.  Similarly, the
``hard-scattering'' contribution is identified with $\hat{H}$, and is
synonymous with ``factorizable'', ``$B$-type'' and
``symmetry-breaking''.  Since $\hat{\zeta}$ involves contributions
from hard gluon exchange in addition to true soft-overlap
contributions from endpoint configurations, dominance of hard-gluon
exchange is not necessarily the same as dominance of $\hat{H}$ over
$\hat{\zeta}$.  The numerical value of $\delta$ provides a useful and
unambiguous means of comparing the implications of different
approaches.

The heavy-quark expansion for exclusive heavy-meson decay
amplitudes yields results such as the well-known symmetry relations
between different form factors, and between different heavy mesons,
which are strictly valid when no other large energy scales are
relevant.  For decays into energetic hadrons, hard-scattering
contributions involving the spectator degrees of freedom in the heavy
meson involve such a new large scale, whose effects may be treated by
the usual approach to hard exclusive processes in the large-recoil
expansion.  The scale separations can be systematically performed
using recently-developed effective field theory techniques, and for
the simplest case of heavy-to-light form factors, relations
(\ref{eq:secondclassSCET}) and (\ref{eq:BDSCET}) give the resulting
modifications to heavy-quark symmetry relations appearing at large
recoil.  The heavy-quark and large-energy scaling laws can be used to
inform extrapolations and parameterizations of the form factors,
e.g. (\ref{eq:newBK}).  Knowledge of the form factors can in turn be
used to disentangle different contributions to more complicated
processes such as $B\to\pi\pi$.

\subsection*{Acknowledgments}
Thanks are due to T.~Becher, S.~Lee and M.~Neubert for 
collaboration on projects
underlying the topics discussed here, and to T.~Becher, A.~Kronfeld 
and M.~Neubert for useful
comments on the manuscript.  
The author acknowledges interesting conversations with M.~Peskin, 
discussions with J.~Dingfelder and L.~Hsu on the experimental analysis, 
and insights from M.~Okamoto and J.~Shigemitsu on the lattice data. 
The hospitality of the TRIUMF theory group and KITP (Santa Barbara) 
is gratefully acknowledged, 
for brief visits where a portion of this work was completed. 
Thanks also to the organizers
of the KITP workshop ``Modern Challenges for Lattice Field Theory'', 
where an early version
of this work was reported.  
Research supported by the Department of Energy under Grant DE-AC02-76SF00515.

\end{document}